\documentclass[sigconf]{acmart}
\acmConference[xxx]{xxxx}{xxxxx}{xxxxx}

\usepackage{soul}
\usepackage{hyperref}
\usepackage{multirow}
\usepackage{fancybox}
\usepackage{enumitem}
\usepackage{graphicx}
\usepackage{balance}
\usepackage{color}
\usepackage{float}
\usepackage{adjustbox}
\usepackage{xcolor}
\usepackage{array}
\usepackage{amsmath}
\usepackage{xspace}
\usepackage{url}
\usepackage{tikz}
\usepackage{caption}
\usepackage{pgfplots}
\usepackage{listings}
\usepackage{color}
\usepackage{graphicx}
\definecolor{dkgreen}{rgb}{0,0.6,0}
\definecolor{gray}{rgb}{0.5,0.5,0.5}
\definecolor{mauve}{rgb}{0.58,0,0.82}
\pgfplotsset{compat=1.16}
\usepackage{pgf-pie}
\usetikzlibrary{pgfplots.statistics,calc}
\usepackage{algorithm}
\usepackage{algpseudocode}
\usepackage{subcaption}
\usepackage{listings}

\usepackage{tcolorbox}

\makeatletter
\newcommand{\mybox}[1]{%
	\setbox0=\hbox{#1}%
	\setlength{\@tempdima}{\dimexpr\wd0+13pt}%
	\begin{tcolorbox}[boxrule=0.5pt, colback=white, arc=4pt,
		left=6pt,right=6pt,top=6pt,bottom=6pt,boxsep=0pt]
		#1
	\end{tcolorbox}
}

\definecolor{codegreen}{rgb}{0,0.6,0}
\definecolor{codegray}{rgb}{0.5,0.5,0.5}
\definecolor{codepurple}{rgb}{0.58,0,0.82}
\definecolor{backcolour}{rgb}{0.95,0.95,0.92}

\lstdefinestyle{mystyle}{
  language=Python,
  aboveskip=3mm,
  showstringspaces=false,
  columns=flexible,
  numbers=none,
  backgroundcolor=\color{backcolour},
  commentstyle=\color{codegreen},
 keywordstyle=\color{magenta},
    numberstyle=\tiny\color{codegray},
    stringstyle=\color{codepurple},
    basicstyle=\small\ttfamily,
    breakatwhitespace=false,         
    breaklines=false,                 
    captionpos=b,                    
    keepspaces=false,                 
    numbersep=5pt,                  
    showspaces=false,                
    showstringspaces=false,
    showtabs=false,                  
    tabsize=2,
    escapeinside=``
}
\definecolor{nima2}{RGB}{1.0, 0.49, 0.0}
\definecolor{songcolor}{RGB}{191,191,191}
\definecolor{nimacolor}{RGB}{0.13, 0.67, 0.8}
\definecolor{aruncolor}{RGB}{51,255,51}

\newcommand{\tool}{\texttt{Orion}}

\AtBeginDocument{%
  \providecommand\BibTeX{{%
    \normalfont B\kern-0.5em{\scshape i\kern-0.25em b}\kern-0.8em\TeX}}}

\setcopyright{acmcopyright}
\copyrightyear{2018}
\acmYear{2018}
\acmDOI{XXXXXXX.XXXXXXX}

\acmBooktitle{Woodstock '18: ACM Symposium on Neural Gaze Detection,
 June 03--05, 2018, Woodstock, NY} 
\acmPrice{15.00}
\acmISBN{978-1-4503-XXXX-X/18/06}

\begin{document}


\title{Security Knowledge-Guided Fuzzing of Deep Learning Libraries}

\author{Nima Shiri Harzevili}
\affiliation{%
  \institution{York University}
  \city{Toronto}
  \country{Canada}}
\email{nshiri@yorku.ca}

\author{Mohammad Mahdi Mohajer}
\affiliation{%
  \institution{York University}
  \city{Toronto}
  \country{Canada}}
\email{mmm98@yorku.ca}

\author{Moshi Wei}
\affiliation{%
  \institution{York University}
  \city{Toronto}
  \country{Canada}}
\email{moshiwei@yorku.ca}

\author{Hung Viet Pham}
\affiliation{%
  \institution{York University}
  \city{Toronto}
  \country{Canada}}
\email{hvpham@yorku.ca}

\author{Song Wang}
\affiliation{%
  \institution{York University}
  \city{Toronto}
  \country{Canada}}
\email{wangsong@yorku.ca}

\begin{abstract}
Recently, many Deep Learning (DL) fuzzers have been proposed for API-level testing of DL libraries. However, they either perform unguided input generation (e.g., not considering the relationship between API arguments when generating inputs) or only support a limited set of corner case test inputs. Furthermore, a substantial number of developer APIs crucial for library development remain untested, as they are typically not well-documented and lack clear usage guidelines, unlike end-user APIs. This makes them a more challenging target for automated testing. 

To fill this gap, we propose a novel fuzzer named {\tool}, which combines guided test input generation and corner case test input generation based on a set of fuzzing heuristic rules constructed from historical data that is known to trigger critical issues in the underlying implementation of DL APIs. To extract the fuzzing heuristic rules, we first conduct an empirical study regarding the root cause analysis of 376 vulnerabilities in two of the most popular DL libraries, i.e., PyTorch and TensorFlow. We then construct the fuzzing heuristic rules based on the root causes of the extracted historical vulnerabilities. Using these fuzzing heuristic rules, {\tool} performs fuzzing to generate corner case test inputs for API-level fuzzing. Additionally, we extend the seed collection of the existing studies to collect test inputs for developer APIs. 

Our evaluation shows that {\tool} reports 135 vulnerabilities on the latest releases of TensorFlow and PyTorch, 76 of which were confirmed by the library developers. Among the 76 confirmed vulnerabilities, 69 are previously unknown, and 7 have already been fixed. The rest are awaiting further confirmation. In terms of end-user APIs, {\tool} was able to detect 31.8\% and 90\% more vulnerabilities on TensorFlow and PyTorch, respectively, compared to the state-of-the-art conventional fuzzer, i.e., DeepRel. When compared to the state-of-the-art LLM-based DL fuzzer, AtlasFuzz, {\tool} detected 13.63\% more vulnerabilities on TensorFlow and 18.42\% more vulnerabilities on PyTorch. Regarding developer APIs, {\tool} stands out by detecting 117\% more vulnerabilities on TensorFlow and 100\% more vulnerabilities on PyTorch compared to the most relevant fuzzer designed for developer APIs, such as FreeFuzz.





\end{abstract}



\maketitle

\section{Introduction}
\label{sec:introduction}


The usage of deep learning (DL) in modern software systems is now widespread and continues to grow. Many successful safety-critical applications have been built on top of DL libraries such as autonomous driving systems~\cite{zhang2018deeproad, zhou2020deepbillboard, garcia2020comprehensive}, healthcare~\cite{zhang2023st, benedetti2023mixed}, and financial services~\cite{fang2023movement}.
As a consequence, there has been a growing concern about vulnerabilities in DL systems. Moreover, as DL applications become new targets for malicious attacks~\cite{harzevili2022characterizing,ji2018model}, reliability and robustness become critical requirements for these applications due to their potential impact on human life~\cite{hong2020artificial, byun2020manifold, barrett2023identifying, carlini2023llm, sun2020automatic, byun2019input, cofer2020run, lutellier2020coconut}.

\begin{figure*}[h]
    \centering
     \includegraphics[scale=1.2]{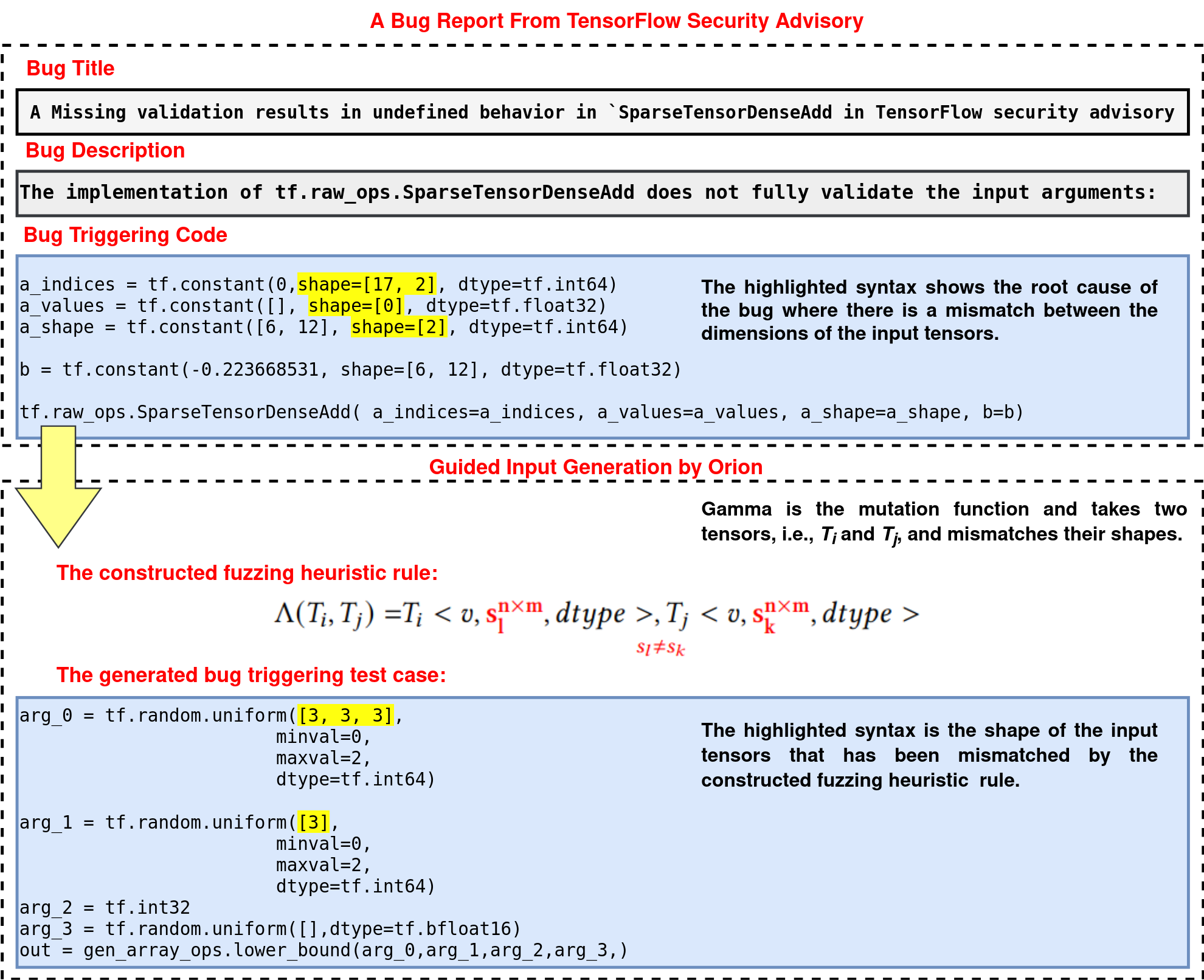}    
     \caption{A motivational example that shows how {\tool} used TensorFlow security advisory to build a fuzzing heuristic rule that models the correlation between input parameters of the API \textit{gen\_array\_ops.lower\_bound} which eventually resulted in the detection of a segmentation fault.}
    \label{fig:check}
\end{figure*}

In recent years, many fuzzers have been proposed for fuzzing DL APIs~\cite{deng2022fuzzing, wei2022free, xie2022docter, deng2023large1, deng2023large2}. Specifically, FreeFuzz~\cite{wei2022free} performed API-level testing via random fuzzing on end-user APIs of TensorFlow and PyTorch collected from three different sources, i.e., API reference documentation, DL models in the wild, and developer tests. DeepRel~\cite{deng2022fuzzing} extended FreeFuzz by random mutation of semantically related APIs in terms of equivalence values and statuses. DocTer~\cite{xie2022docter} performed fuzz testing via automatic extraction of constraints from the API reference documentation of three DL libraries (i.e., TensorFlow, PyTorch, and MXNet). TitanFuzz~\cite{deng2023large1} leveraged the power of Large Language Models (LLMs)~\cite{first2023baldur, wei2023copiloting, zhang2023multilingual} to generate inputs tailored for testing DL APIs. AtlasFuzz~\cite{deng2023large2} extended TitanFuzz by including unusual programs mined from open source by LLMs to guide the test input generation. Although existing DL fuzzers have shown great potential in finding real-world DL vulnerabilities, they suffer from the following limitations: 

\noindent \textbf{Challenge 1:} \noindent \textbf{Unguided Input Generation}. 
Existing DL fuzzers~\cite{wei2022free, xie2022docter, deng2022fuzzing, deng2023large1, deng2023large2} mainly treat the parameters of DL APIs independent in API level fuzz testing. However, during our empirical study, we find that some vulnerabilities require a special combination of different input arguments, e.g., \texttt{mismatch between input arguments} of DL APIs is one of the major root causes of reported security vulnerabilities. Figure~\ref{fig:check} shows an example vulnerability\footnote{https://github.com/tensorflow/tensorflow/security/advisories/GHSA-rc9w-5c64-9vqq} exposed by \textit{mismatch between dimensions of the input tensors} in earlier releases of TensorFlow. The description of the vulnerability report states that the implementation of \textit{tf.raw\_ops.SparseTensorDenseAdd} lacks validation on the input arguments, resulting in undefined behavior. According to the API reference documentation of \textit{tf.raw\_ops.SparseTensorDenseAdd}, the first dimension of \textit{a\_values} is expected to match with the first dimension of \textit{a\_indices}, and the first dimension of \textit{a\_shape} should align with the second dimension of \textit{a\_indices}. This complicated relationship imposes a challenge for existing fuzzers, as they typically lack the capability to explore and discover such nuanced correlations among the input arguments of DL APIs. As a result, triggering vulnerabilities exposed by \textit{guided input generation} becomes practically implausible within the scope of conventional and LLM-based fuzzing techniques.



\noindent \textbf{Solution:} \textbf{Guided Input Generation using Historical Rules.}
To address \textbf{Challenge 1}, we propose guiding the generation of test inputs for fuzzing DL APIs using heuristic rules derived from historical vulnerabilities triggered by specific DL API argument patterns. \textit{Our key insight on the guided input generation of test inputs is that security vulnerabilities in DL libraries often manifest when APIs encounter a mismatch between input arguments.} Figure~\ref{fig:check} (the lower box) provides an example fuzzing heuristic rule and the corresponding generated test case illustrating our approach for solving the aforementioned problem by focusing on the \texttt{mismatch between input arguments}. Initially, we conduct a root cause analysis by summarizing a set of historical vulnerabilities attributed to~\texttt{mismatches between input arguments of DL APIs} (refer to Section~\ref{sec:empiricalStudy}). Leveraging the insights from this root cause analysis, we formulate a set of fuzzing heuristics rules designed to produce targeted test inputs capable of exposing such vulnerabilities (refer to Table~\ref{tbl:mismatchRules} for a detailed explanation of these fuzzing heuristics rules). As shown in Figure~\ref{fig:check}, the constructed rule, in the form of a mutator function alters the dimensions of input tensors, ultimately triggering a segmentation fault. More specifically, the generated test case features two input tensors with conflicting dimensions: the first tensor possesses a \texttt{(3D)} shape, while the second tensor has a \texttt{(2D)} shape. The generated test inputs violate the constraints outlined in the API reference documentation for \textit{tf.raw\_ops.lower\_bound}, which specifies that both \textit{sorted\_inputs} and \textit{values} should be \texttt{(2D)} tensors.

\noindent \textbf{Challenge 2: Lack of Support for Corner Case Generation.}
Existing conventional DL fuzzers at the API level~\cite{wei2022free, deng2022fuzzing, xie2022docter, deng2023large1, deng2023large2} suffer from two primary limitations when it comes to generating corner case test inputs. First, their heuristics rules for corner case test input generation are \textit{random} and \textit{unguided}. Random values may not accurately replicate real-world vulnerabilities or tainted inputs provided by external attackers. API calls often rely on specific data formats, constraints, or patterns that random values may not adhere to. This can result in false positives or missed detection of critical vulnerabilities. This can lead to false positives or missing the detection of critical vulnerabilities. For instance, consider a documented issue within the TensorFlow GitHub repository\footnote{https://github.com/tensorflow/tensorflow/issues/51908}. In this case, a TensorFlow user reported a crash in the \textit{tf.pad} API when called with \texttt{large padding sizes}. Our proposed fuzzer, {\tool}, successfully identified this security vulnerability, while it went unnoticed by other studied DL fuzzers. The reason behind this lies in their approach of generating arbitrary test inputs for DL APIs, rendering them essentially incapable of detecting vulnerabilities that manifest only under extreme corner-case conditions. As for LLM-based DL fuzzers like TitanFuzz and AtlasFuzz, they primarily focus on addressing logical bugs by modeling context information and DL API usage call sequences, rather than detecting security vulnerabilities through corner case test input generation. 


\noindent \textbf{Solution: Guided Corner Case Generation using Historical Rules.}
To effectively address the challenge of limited corner case test input generation, we analyze historical security vulnerability data to identify fuzzing heuristic rules pertaining to the types of inputs, corner case conditions, or unusual scenarios that have historically resulted in software security vulnerabilities. These corner case fuzzing heuristic rules (as presented in Table \ref{tbl:mismatchRules}) serve as guidance for {\tool}'s corner case generator, instructing it on how to intelligently generate or mutate test inputs. \textit{What makes these constructed fuzzing heuristic rules particularly valuable, compared to the traditional and LLM-based DL fuzzing techniques, is their ability to accurately mimic real-world corner cases that have historically exposed software vulnerabilities.}

\noindent \textbf{Challenge 3: Lack of Testing Developer APIs}. Current conventional and LLM-based DL fuzzers \cite{wei2022free, deng2022fuzzing, xie2022docter, deng2023large1, deng2023large2} predominantly concentrate their testing efforts on end-user DL APIs. This focus arises from the fact that DL libraries often lack comprehensive documentation and usage guidelines for developer APIs, in contrast to the well-documented nature of end-user APIs.  

\noindent \textbf{Solution: Modeling Developer API Context Information}.
To overcome the challenge, we developed a lightweight static analyzer and collected developer API context information heuristically. The core idea is to focus on internal Python modules within the DL libraries where all developer APIs are located which act as an intermediate between Python client APIs, i.e., end-user APIs, and the backend implementation of DL libraries. 


Our assessment demonstrates that {\tool} has identified a total of 135 vulnerabilities 
on the latest releases of TensorFlow (2.13.0) and PyTorch (1.13.1), and 76 of them have been verified by library developers. Out of these 76 confirmed vulnerabilities, 69 were previously unknown, and 7 have already been addressed and resolved. The remaining bugs are pending further confirmation. Specifically, \textbf{Regarding end-user APIs}, {\tool} uncovers 31.8\% and 90\% additional vulnerabilities on TensorFlow and PyTorch compared to the leading conventional DL fuzzer, i.e., DeepRel. Compared to the cutting-edge LLM-based fuzzer, i.e., AtlasFuzz, {\tool} could detect 13.63\% and 18.42\% more vulnerabilities on TensorFlow and PyTorch respectively. \textbf{In terms of developer APIs}, {\tool} could outperform FreeFuzz (the only applicable baseline) by detecting 117\% and 100\% more vulnerabilities on TensorFlow and PyTorch respectively. This paper makes the following contributions: 

\begin{itemize}

\item We present a simple yet effective approach to API-level testing of deep learning libraries. To the best of our knowledge, this is the first study to leverage historical vulnerability data for the creation of fuzzing heuristic rules. 

\item We characterize and categorize a set of fuzzing heuristic rules based on an empirical study of 376 security records extracted from the TensorFlow security advisory\footnote{https://github.com/tensorflow/tensorflow/security/advisories.} and GitHub repository of PyTorch.

\item We design and develop a DL fuzzer, {\tool}, that utilizes the fuzzing heuristics rules to guide its input generation. {\tool} also instruments both end-user and developer APIs specifically designed to test downstream components of DL libraries. 

\item {\tool} can find 135 security vulnerabilities of which 76 of them are already confirmed by library developers. Among the confirmed security vulnerabilities, 69 of them are new and 7 of them are already fixed. The rest of the reported security vulnerabilities are pending and awaiting for further confirmation.

\end{itemize}

\section{Empirical Analysis of History Security Vulnerabilities}
\label{sec:empiricalStudy}

To characterize and gain insights into fuzzing heuristic rules, we conducted an empirical study on 376 historical security vulnerabilities in PyTorch and TensorFlow to understand their root causes. In this section, we explain our approaches to collecting and analyzing these records, as well as extracting the fuzzing heuristic rules that {\tool} employs to guide its fuzzing operations.




\subsection{Data Collection}
In this work, we gathered historical security vulnerability data from two distinct sources for TensorFlow and PyTorch: issues in the GitHub repository for PyTorch and the TensorFlow security advisory\footnote{https://github.com/tensorflow/tensorflow/security/advisories}. The reason for collecting security records from GitHub for PyTorch is that, at the time of writing this paper, there were only four reported security vulnerabilities in its CVE repository\footnote{https://cve.mitre.org/cgi-bin/cvekey.cgi?keyword=pytorch}. For TensorFlow, we meticulously reviewed all 407 reports available at the time of writing this paper on its security advisory page. 


\subsubsection{Automated collection of real-world security vulnerabilities}
For the automatic collection of issues from PyTorch, we iterated through all available issues in its repository. We employed keyword-based matching approaches, similar to those used in~\cite{song2021, tian2012identifying, zhou2017automated}, to automatically filter out irrelevant issues and gather those related to security vulnerabilities. Our chosen keywords included:

\noindent \textbf{Numerical and Memory-related vulnerabilities}: \textit{buffer overflow}, \textit{integer overflow}, \textit{integer underflow}, \textit{heap buffer overflow}, \textit{stack overflow}, and \textit{null pointer dereference}\footnote{To save space, we have not incorporated all security-related terms in the manuscript. However, a comprehensive list of all keywords can be found in the GitHub repository associated with this paper.}. \noindent \textbf{Logical vulnerabilities}: \textit{wrong result}, \textit{unexpected output}, \textit{incorrect calculation}, \textit{inconsistent behavior}, \textit{unexpected behavior}, \textit{incorrect logic}, \textit{wrong calculation}. \noindent \textbf{Performance vulnerabilities}: \textit{slow}, \textit{high CPU usage}, \textit{high memory usage}, \textit{poor performance}, \textit{slow response time}, \textit{performance bottleneck}, \textit{performance optimization}, \textit{resource usage}, \textit{race condition}, \textit{memory leak}. 
As a result, we collected 1,739 vulnerability-related issues from the GitHub repository of PyTorch. However, our manual analysis revealed that such automated approaches produce numerous false positives. This occurs because not all issues in the PyTorch GitHub repository pertain to actual software vulnerabilities. Many unrelated issues involve CI infrastructure, documentation, and feature requests. In the following section, we will discuss our manual filtering approach to remove these unrelated issues.

\subsubsection{Manual filtering}

We manually examined the collected data to get the final list of vulnerability-related issues for PyTorch. Specifically, for each issue, we carefully reviewed the issue title, description, discussions, log messages, etc. We exclude 1) Vulnerabilities specific to certain platforms, such as Windows, Android, or IOS; 2) Build and configuration issues; 3) Bugs arising from external libraries such as torchvision or torchaudio; 4) Bugs that do not require input parameters to be triggered. 
After applying the exclusion and inclusion criteria, we identified a total of 98 issues related to security vulnerabilities in PyTorch.

In total, our root cause analysis encompassed 376 DL security vulnerability records, with 278 records sourced from TensorFlow and 98 records obtained from PyTorch.

\begin{table}[]
    \centering
    \caption{The notations we used in this paper.}
    \vspace{-0.1in}
    \resizebox{1\columnwidth}{!}{
    \begin{tabular}{ll}
    \toprule
    Notation & Explanation \\
    \midrule
    \small $\Lambda$ & The mutation function. \\
    \small $T_j<v,s_{n\times m},dtype>$ & A triplet which defines a tensor. \\
    \small $v$  & The value of the tensor subject to  $v \in \{\mathbb{Z}, \mathbb{R}, \mathbb{C}\}$.     \\
    \small $s_{n\times m}$  & Is the shape of the tensor denoted as a 2D tensor.                                     \\
    \small $dtype$   & Denotes the tensor data types subject to $dtype \in DT$. \\
    \small  $n,m,l,i,j,k$ & A set of indices subject to $n,m,l,i,j,k \in \mathbb{Z} \quad \land  \neq \mathbb{C} |  \mathbb{R}$.\\
    \small $case_x$ & Large/Zero value corner case generator subject to $case_x \in \mathbb{R}| case_x \in \mathbb{Z}$.  \\
    \small $case_n$ & Negative value corner case generator subject to $case_n \in \mathbb{R}| case_n \in \mathbb{Z}$. \\
    \small $case_{nan}$ & NaN corner case generator.\\
    \small $case_{none}$ & Python $None$ corner case generator. \\
    \small $case_{mt}$ & Empty value corner case generator. \\
    \small $case_{noa}$ & Non-ASCII character string corner case generator. \\
    \small $arg$ & A dummy argument which can take any type. \\
    \small $L$ & denotes any Python list or tuple.\\
    \bottomrule
    \end{tabular}
    \label{tbl:notations}}
\end{table}

\begin{table}[]
\caption{Summary of fuzzing heuristic rules.} 
\resizebox{1\columnwidth}{!}{%
\begin{tabular}{p{0.1cm}ll}
\toprule
\textbf{No~}  &~\textbf{Guided Input Generation Rules}  &      \textbf{Rule Notation}          \\
\midrule 
1& Tensors Shape Mismatch &$\Lambda(T_i, T_j) = \underset{\textcolor{red}{s_l\neq s_k}}{T_i<v, \textcolor{red}{\mathbf{s_l^{n\times m}}} ,dtype>,T_j<v,\textcolor{red}{\mathbf{s_k^{n\times m}}},dtype>}$  \\
\\[0.1ex]
2&Tensor Dimension Mismatch& $\Lambda(T_l, arg_i) = \underset{\textcolor{red}{arg_i \neq s_{n\times m}}}{\{T_1<v,\textcolor{red}{\mathbf{s_{n\times m}}},dtype>,\textcolor{red}{\mathbf{\underset{arg_i \in \mathbb{Z}}{arg_i}}}\}}$ \\
\\[0.1ex]
3&Tensor List-Indices Mismatch & $\Lambda(T_l, L_i) = \underset{\textcolor{red}{|L_i|\neq|n\times m|}}{T_l<v,\textcolor{red}{\mathbf{s_{n\times m}}},dtype>}, \textcolor{red}{\mathbf{L_i}}=\{k_1,k_2,...,k_j\}$   \\
 \\[0.1ex]
4&List Indices Elements Mismatch & $\Lambda(T_l,L_i) = \underset{ \textcolor{red}{\mathbf{L_{i,j}\neq n | L_{i,j} \neq m}}}{T_l<v,\textcolor{red}{\mathbf{s_{n\times m}}},dtype>}, L_i=\{k_1,k_2,...,\textcolor{red}{\mathbf{k_j}}\}$  \\
 \\[0.1ex]
5&List Indices Length Mismatch &$\Lambda(L_i, L_j) = \underset{\textcolor{red}{\mathbf{|L_i|}} \neq \textcolor{red}{\mathbf{|L_j|}}}{\textcolor{red}{\mathbf{L_i}}=\{x_1,x_2,...,x_n\},\textcolor{red}{\mathbf{L_j}}=\{a_1,a_2,...,a_k}\}$     \\
\hline \hline
&\textbf{Corner Case Input Generation Rules} &  \textbf{Rule Notation}          \\
\hline 
6&Tensor Corner Case Generator Type 1&$\Lambda (T_i) = {T_i<\textcolor{red}{\mathbf{\{case_x|case_n|case_{nan}\}}},s_{n\times m},dtype>}$  \\
\\[0.1ex]
7&Tensor Corner Case Generator Type 2&$\Lambda (T_i) = {T_i<v,s_{\textcolor{red}{\mathbf{\{case_x|case_n\}}}\times m},dtype>}$  \\
\\[0.1ex]
8&Tensor Corner Case Generator Type 3 &$\Lambda (T_i) = {T_i<v,s_{n \times \textcolor{red}{\mathbf{\{case_x|case_n\}}}},dtype>}$ \\
\\[0.1ex]
9~ &~Scalar Tensor Corner Case Generator &$\Lambda(T_i) = \underset{n=1\land m=1}{T_i<v, s^{n\times m},dtype>}$ \\
\\[0.1ex]
{10~} &~Non-Scalar Tensor Corner Case Generator &$\Lambda(T_i^{s=1}) = T_i<v, s_{n\times m},dtype>$ \\
\\[0.1ex]
11 &~Preemptive Corner Case Generator Type 1 &$\Lambda (\underset{arg_i \in \mathbb{Z|R}}{arg_i}) = \textcolor{red}{case_x|case_n|case_{nan}|case_{none}|case_{mt}|case_{noa}} $ \\ 
\\[0.1ex]
12~ &~Preemptive Corner Case Generator Type 2&$\Lambda (\textcolor{red}{\mathbf{\underset{arg_i \in Boolean}{arg_i}}}) = \textcolor{red}{\mathbf{\{case_x|case_n\}}}$ \\ 
\\[0.1ex]
13~ &~Preemptive Corner Case Generator Type 3 &$\Lambda (\textcolor{red}{\mathbf{\underset{arg_i \in String}{arg_i}}}) = \textcolor{red}{\mathbf{\{case_{mt}|case_{noa}\}}}$ \\
14~ &~List/Tuple Corner Case Generator &$\Lambda (L_i) = \underset{k_j \in \textcolor{red}{\mathbf{\{case_x|case_n|case_{nan}|case_{none}|case_{mt}|case_{noa}\}}}}{L_i = \{k_1, k_2, \ldots, k_j\}}$ \\
\\[0.1ex]
\bottomrule
\end{tabular}}
\label{tbl:mismatchRules}
\end{table}

\subsection{Construction of Fuzzing Heuristics Rules}
\label{sec2.2}

To construct fuzzing heuristic rules, we initially conducted manual analyses of the collected vulnerability records to explore potential input patterns contributing to these vulnerabilities. For CVE records, we first thoroughly examined the description of reported vulnerabilities, as well as the provided links to the TensorFlow security advisory. Subsequently, we reviewed the vulnerability description, minimum reproducing example, and the link to the commit that patched the security issue. Regarding commits, we focused on code changes to identify the root cause of the issue in the backend implementation. In the case of PyTorch security records, our review encompassed issue descriptions, discussions, related issues, and reproducing examples, all aimed at identifying the factors contributing to vulnerabilities. Specifically, we extracted the following factors from each record:

\noindent \textbf{Root cause}. This factor represents the root cause of vulnerabilities in DL libraries, which is crucial when constructing fuzzing heuristic rules. We extracted 33 unique root causes\footnote{Due to space limitations, the details of the summarized root causes are available in the supplementary documentation at \cite{ourdata}} from 378 security records for both TensorFlow and PyTorch. 

\noindent \textbf{Reproducing example}. We checked for the presence of stand-alone code examples to reproduce the vulnerability. Reproducing examples helps us understand which input specifications to DL APIs trigger vulnerabilities, aiding in the implementation of fuzzing heuristic rules.

\noindent \textbf{Vulnerable Parameter and and its Type}. We also extracted information about the vulnerable parameter in the API input specification and its type. Collecting the type of vulnerable parameter is important because each parameter type has its own weaknesses in terms of fuzz testing. For instance, concerning tensors, two main vulnerable components are typically identified: tensor values and tensor shapes.

Ultimately, we categorized various fuzzing heuristic rules based on the 33 unique root causes, and the details are presented in Table~\ref{tbl:mismatchRules}. It's important to note that the definition of fuzzing heuristic rules is consistent for both libraries, with the only difference being their implementation, which is specific to each library. In Table~\ref{tbl:mismatchRules}, we divided the fuzzing heuristic rules into two sections: \textit{Guided Input Generation Rules} and \textit{Corner Case Input Generation Rules}. The first section consists of fuzzing heuristic rules that involve mismatches in the input arguments of DL APIs. For example, in the first rule, labeled \textit{Tensors Shape Mismatch}, we employ a general indicator function denoted as $\Lambda$, which acts as a mutator. This function takes two input tensors, $T_i$ and $T_j$, and performs a shape mismatch operation sequentially. To better comprehend this process, let's consider the manipulation of two tensors, $T_i$, within the parameter space of the API. The sequence of operations begins by altering the shape of the first tensor, $T_i$. This shape modification can be achieved through either \texttt{Rank Reduction} or \texttt{Rank Expansion}, with the choice between these two operations being random. {\tool} keeps track of the previous operation in a temporary variable for reference. Now, as the shape mutation of the second parameter begins, the process retrieves the previously recorded operation from the temporary variable. It then performs the opposite operation compared to what was executed on the first parameter. In other words, if the initial operation on the first parameter was \texttt{Shape Reduction}, the new operation applied to the second parameter is \texttt{Shape Expansion}, and vice versa. This rule ensures that the shape mismatch is consistently executed as intended

In the second part, we have organized these corner case input generation rules based on the type of input parameters utilized by DL APIs. To illustrate, let's consider the initial set of three rules (from row 1 to row 3 in Table~\ref{tbl:mismatchRules}), which primarily target the tensor data type. For example, the first rule in this category deals with the manipulation of input tensor values. Within this rule, there are three distinct corner case test input generators (highlighted in red): $case_x$: This represents the corner case generator for large or zero test inputs. $case_n$: Indicates the generator for negative value corner case test inputs. $case_{nan}$: Indicates the corner case test input generator for \texttt{NaN} test inputs.

\begin{figure*}[t!]
    \centering
     \resizebox{\textwidth}{!}{{\includegraphics{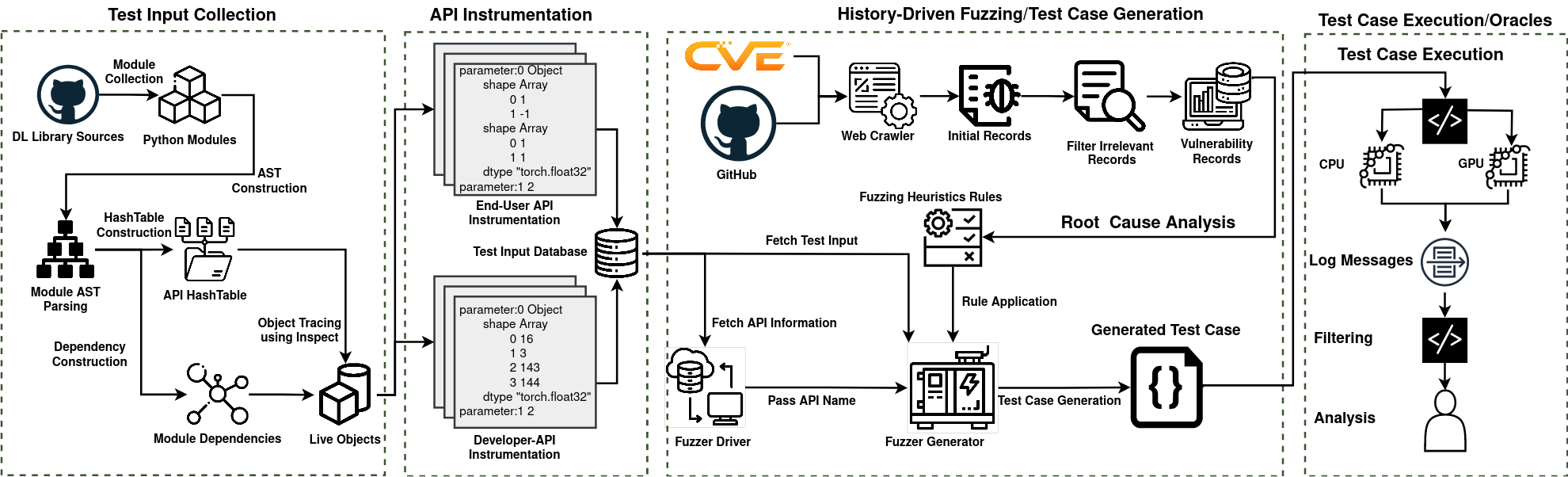}}}    
     \caption{Overview of our proposed {\tool}.}
     
    \label{framework}
\end{figure*}

{\tool} utilizes the fuzzing heuristic rules summarized in Table~\ref{tbl:mismatchRules} to instruct its fuzzer generator on how to perform mutations on the test inputs gathered from various sources. This process is illustrated in our framework, as shown in Figure~\ref{framework}.

\definecolor{dkgreen}{rgb}{0,0.6,0}
\definecolor{gray}{rgb}{0.5,0.5,0.5}
\definecolor{mauve}{rgb}{0.58,0,0.82}

\definecolor{dkgreen}{rgb}{0,0.6,0}
\definecolor{gray}{rgb}{0.5,0.5,0.5}
\definecolor{mauve}{rgb}{0.58,0,0.82}

\section{Framework}
\label{sec:framework}


\noindent \textbf{Test Input Collection (Stage 1):} In this initial stage, API test inputs are gathered from various sources, including library documentation, developer tests, and publicly available repositories on GitHub that utilize TensorFlow and PyTorch APIs. Further details on this data collection process can be found in subsection~\ref{sec:4.2}.

\noindent \textbf{API Instrumentation (Stage 2):} In the second stage, dynamic instrumentation is employed to trace execution details for each API invocation. This information encompasses parameter values and types. The collected data is then used to create a type space, API value space, and argument value space, which are crucial for the subsequent stages. More information on dynamic instrumentation can be found in subsection~\ref{sec:4.2}.

\noindent \textbf{History-Driven Fuzzing/Test Case Generation (Stage 3):} The third stage introduces our history-driven fuzzer, as described in subsection~\ref{sec:3.3}. Utilizing fuzzing heuristic rules, {\tool} conducts fuzzing on the test inputs retrieved from its database and generates the corresponding test cases.

\noindent \textbf{Test Execution and Oracles (Stage 4):} The final stage involves test execution and oracles, detailed in subsection~\ref{sec:4.4}. Test cases generated in the previous stage are executed, and their log messages are filtered based on predefined oracles. This stage includes running the test cases under two different settings: CPU and GPU. Log messages are parsed according to the specified oracles. Subsequently, manual analysis is performed to eliminate irrelevant test cases, such as those resulting from syntax errors or user errors.

\subsection{Test Input Collection and API Instrumentation}
\label{sec:4.2}
 
Following existing work~\cite{wei2022free, deng2022fuzzing}, {\tool} performs test input collection from three sources, i.e., API reference documentation, publicly available repositories, and developer tests. {\tool} then performs API instrumentation to collect various dynamic execution information of both end-user and developer APIs (i.e., the type and value for each parameter). In this work, we followed FreeFuzz~\cite{wei2022free} to instrument end-user APIs for collecting their dynamic execution information. A significant challenge in automated fuzz testing of DL libraries arises from the fact that many developer APIs that are essential for library development, are often overlooked. This is primarily due to their lack of comprehensive documentation and clear usage guidelines, which contrast with end-user APIs. In order to tackle the problem, we extended FreeFuzz instrumentation to make it compatible with developer APIs.

To address this challenge, we devised a solution, shown as the first box in Figure~\ref{framework}, in the form of a lightweight static analyzer, leveraging the capabilities of the Python AST module\footnote{https://docs.python.org/3/library/ast.html}. This analyzer plays a pivotal role in the systematic collection of vital data pertaining to developer APIs within the TensorFlow library. Firstly, {\tool} clones the source code of DL libraries from GitHub and extracts all relevant Python files within the Python directory of the source codes. These modules act as intermediaries between the backend, where extensive computations occur, and the client code accessible to end-users. Next, for each Python module file, we construct a corresponding Abstract Syntax Tree (AST). We make use of the AST visitor functionality provided by the Python AST module to systematically extract the names of developer APIs present within each module. The extracted developer API names are organized into categories based on their respective parent modules. Simultaneously, we construct dependency statements for each developer's API name. To ensure the usability of the developer API names, we utilize the Python inspect module\footnote{https://docs.python.org/3/library/inspect.html}. This module allows us to access live module object information associated with each developer's API name, a crucial step in facilitating test input collection. Finally, we pass the live objects for API instrumentation, covering both end-user and developer APIs. 


\subsection{History-Driven Test Case Generation}
\label{sec:3.3}

Before we delve into explaining the process of fuzzing, we define the following notations:
\begin{itemize}
    \item The API name, denoted as $apiName$.
    \item The randomly fetched test input, denoted as $\{p_1, p_2, ...,p_j\}$, where $j$ indicates the total number of parameters in the parameter space of the test input corresponding to $apiName$.
    \item $numIter$, representing the total number of times the API is tested.
    \item A collection of fuzzing heuristic rules stored as a set of lookup tables, represented as ${L_{0}(t), L_{1}(t),..., L_{n-1}(t)}$, where:
    \begin{itemize}
        \item $t$ corresponds to the type of current argument under test.
        \item $n$ signifies the total number of lookup tables. The lookup tables store fuzzing heuristics rules for each parameter type.
    \end{itemize}
\end{itemize}

\noindent \textbf{Fuzzer Driver:}
In the initial step, the fuzzer driver retrieves the list of all available DL APIs collected for fuzzing. It then iterates through this list, one API at a time. For each specific API, denoted as $apiName$, the driver passes this API name to the \textit{Fuzzer Generator} component.

\noindent \textbf{Fuzzer Generator:}
In this step, the fuzzer generator receives the API name from the fuzzer driver and retrieves a random test input for the API $apiName$ from the test input database. Subsequently, the generator iterates through all the parameters in the test input. For the current parameter under test, denoted as $p_j$, the generator identifies its type.Once the parameter type ($t$) is determined, the generator extracts a list of available fuzzing heuristic rules specific to that parameter type. Next, {\tool} iterates through all the available fuzzing heuristic rules in the current lookup table and applies each rule, denoted as $r_i$, to the current parameter $p_j$ using its mutation function. To illustrate, if $p_j$ represents a tensor, the generator systematically applies all rules specifically designed for tensors. This iterative process continues until all rules have been applied, ensuring a comprehensive exploration of the API's parameter space and corresponding input values (argument space).

\noindent \textbf{Test Case Generator:} 
During this phase of the fuzzer generator, after the test input has undergone fuzzing based on the extracted fuzzing heuristic rules, the resulting mutated test input is transformed into a Python test case. This transformation involves the inclusion of all necessary information into the test case. This information encompasses error-handling statements, statements for both CPU and GPU computations, and alignment of the test case with specific hardware configurations and computational requirements necessary for accurate execution. Once the test case is generated, the fuzzer generator submits it for execution.

\subsection{Test Case Execution}
\label{sec:4.4}
In this stage, the generated test case is executed to identify potential security vulnerabilities. The log message of the generated test case undergoes automated parsing to eliminate irrelevant execution logs, such as syntax errors and timed-out cases. To ensure the accuracy of our findings and minimize potential bias in the results, we also perform a manual examination and analysis of the output from the executed test cases. During this analysis, we consider the following oracles:

\noindent \textbf{Crash Oracle:}
A \textit{Crash} is defined as an event where the executed test case either halts the Python interpreter or triggers a runtime error. These issues can sometimes be induced by invalid inputs generated during the fuzzing process. To mitigate false alarms automatically, we developed a set of scripts that utilize heuristics to identify and filter out crash incidents that lead to specific exceptions, such as \textit{ValueError} or \textit{InvalidArgumentError}, across all backends.

\noindent \textbf{Differential testing:}
This oracle operates by executing the generated test case under two distinct configurations: one on CPU and the other on GPU. For each configuration, it produces two separate results using identical test inputs. We then undertake manual analysis of these outputs, meticulously examining them for any discrepancies or inconsistencies reported by the API under test.
\section{Experimental Setup}
\label{sec:setup}
\subsection{Research Questions}
To evaluate the performance of {\tool}, we design experiments to answer the following research questions:

\begin{itemize}
    \item \textbf{RQ1}: What is the performance of Orion compared to the four baselines?
    \begin{itemize}
        \item \textbf{RQ1.1}: What is the API coverage rate of Orion compared to the existing DL fuzzers?
        \item \textbf{RQ1.2}: What is the performance of Orion compared to traditional DL fuzzers?
        \item \textbf{RQ1.3} What is the performance of Orion compared to LLM-based DL fuzzers?
    \end{itemize}
    \item \textbf{RQ2}: What is the performance of {\tool} in detecting new vulnerabilities?
    \item \textbf{RQ3}: What is the contribution of {\tool}'s fuzzing heuristic rules in detecting vulnerabilities?
\end{itemize}

\subsection{Subject DL libraries}
We have selected two widely recognized and extensively used DL libraries, namely TensorFlow and PyTorch, as our primary experimental subjects. For RQ1 and in order to compare {\tool} versus traditional DL fuzzers, we have employed earlier releases of TensorFlow (specifically, releases 2.3.0 and 2.4.0) and PyTorch (releases 1.7.0 and 1.8.0) as the subjects of our evaluation. Regarding comparison with LLM-based fuzzers, we use their reported releases for TensorFlow and PyTorch namely releases 2.10.0 and 1.12.0, respectively. For RQ2 RQ3, we have utilized the most recent releases of TensorFlow and PyTorch, namely releases 2.13.0 and 1.13.1, respectively. 

\subsection{Testing Environment}

Our machine is equipped with Intel(R) Core(TM) i7-10700F CPU @ 2.90GHz, NVIDIA GTX 1660 Ti GPU, 16GB of RAM, and Ubuntu 22.04. We run {\tool} and the baseline fuzzers on separate conda environments. For each release, we created an isolated conda environment and installed the all required dependencies required by each fuzzer. In order to manage our system resources and make the comparison fair, we only run one tool at a time. We also run our proposed fuzzer {\tool} \textbf{1000} times for each API.

\subsection{Baseline Approaches}
\label{sec:4.3}
In this paper, we select five state-of-the-art fuzzing tools as baseline techniques, three conventional tools, and two LLM-based tools.


\noindent \textbf{FreeFuzz}~\cite{wei2022free}: is a DL fuzzer that performs fuzz testing on TensorFlow and PyTorch libraries for bug detection. We reuse the replication package of FreeFuzz for comparison and adopt the recommended configurations in our experiments. We configured FreeFuzz to be executed with \textbf{1000} iterations for each API. We configured it to perform value and type mutation. 

\noindent \textbf{DeepRel~\cite{deng2022fuzzing}:} extends FreeFuzz by using test inputs from one API to test other related APIs that share similar input parameters. We carefully analyzed the DeepRel paper and found that its best configuration is 1 iteration with $top\_k=5$. We also run DeepRel \textbf{1000} times for each API. 

\noindent \textbf{DocTer}~\cite{xie2022docter}: is a documentation-driven fuzz testing framework that generates inputs for DL APIs based on specifications mined from API reference documentation. We run the publicly available DocTer implementation shared by the authors in our comparison experiments with the suggested configurations. DocTer's execution involved \textbf{1000} iterations for each API. We carefully studied its paper and realized that conforming inputs are the best setting for bug detection. \footnote{Please note that there is another setting called violating input, though, the number of detected bugs by Conforming inputs is higher than violating inputs.}

\noindent \textbf{TitanFuzz}~\cite{deng2023large1}: is a cutting-edge DL fuzzer, that harnesses the power of Large Language Models (LLMs) for conducting API-level fuzzing on TensorFlow and PyTorch. TitanFuzz uses four different groups of mutation operators for evolutionary fuzzing including argument, suffix, prefix, and method.

\noindent \textbf{AtlasFuzz}~\cite{deng2023large2}: is the extension of TitanFuzz where it uses historical bug data collected from the open source to guide their fuzzier generator with large language models. Note that, different from {\tool}, it uses historical bug data to generate unusual programs to expose critical bugs on TensorFlow and PyTorch releases.

\noindent \textbf{Comparison Settings for TitanFuzz and AtlasFuzz}
In this paper, we leverage the reported vulnerabilities identified in TitanFuzz and AtlasFuzz to address RQ1. The rationale lies in the inherent incompatibilities that arise between the specific DL releases and their corresponding CUDA configurations. To illustrate this point, consider TensorFlow 2.3.0, which mandates the use of CUDA release 10.1, as specified in the official TensorFlow documentation\footnote{https://www.tensorflow.org/install/source}. Conversely, TitanFuzz and AtlasFuzz rely on PyTorch 1.12.1 to execute their model, a requirement documented in their replication package where Torch-1.12.1 is specified. However, PyTorch 1.12.1 necessitates CUDA 10.2 for GPU support, as detailed in the PyTorch release history\footnote{https://pytorch.org/get-started/previous-versions/}. Attempting to run TitanFuzz and AtlasFuzz on TensorFlow 2.3.0, for instance, leads to a \textit{RuntimeError: CUDA error: no kernel image is available for execution on the device}. We observe similar incompatibility issues on PyTorch releases.

\subsection{Evaluation criteria}
\label{sec:4.5}
Following existing work~\cite{xie2022docter,wei2022free}, we use the following evaluation criteria to evaluate the performance of {\tool} and the baselines:

\noindent \textbf{Number of covered APIs}.
The number of covered APIs is a good indicator of how effective a fuzzer can be. The more APIs are covered, the more vulnerabilities will be discovered and fixed. 

\noindent \textbf{Number of vulnerabilities in earlier releases}.
We also apply {\tool} on the earlier releases of TensorFlow and PyTorch including 2.3.0, 2.4.0, 1.7.0, and 1.8.0 respectively to compare its performance with baseline approaches, i.e., FreeFuzz, Docter, and DeepRel. Specifically, we use detection rate and fixed rate as two metrics for the comparison. The detection rate indicates the detection performance of the fuzzer while the fixed rate indicates how many vulnerabilities are fixed in the latest releases for each library. We consider the release 2.13.0 and 1.13.1 as the latest releases for TensorFlow and PyTorch respectively.  

\noindent \textbf{Number of new detected vulnerabilities}.
The primary objective of this study is to uncover new vulnerabilities in the latest releases of TensorFlow and PyTorch, thus we also report the number of new vulnerabilities detected for the studied DL libraries. The new releases include 2.10.0, 2.11.0, and 2.13.0 for TensorFlow and 1.12.0 and 1.13.1 for PyTorch. 

\noindent \textbf{Contribution of Fuzzing heuristics rules}.
We also measure the contribution of the fuzzing heuristic rules introduced in this paper in detecting new vulnerabilities on the latest releases of TensorFlow and PyTorch.

\section{Result Analysis}
\label{sec:results}

\subsection{RQ1: Comparison with Baselines}
To evaluate the performance of {\tool} against the five baselines, we compare the number of covered APIs by each tool across TensorFlow and PyTorch, the detection performance of {\tool} against the traditional DL fuzzers, and the detection effectiveness of {\tool} against LLM-based DL fuzzers. 

\subsubsection{Approach} 

\noindent \textbf{RQ1.1: API coverage}
Regarding TensorFlow in order to be able to run the test cases effectively and perform instrumentation on developer APIs, we build it from source by the default configuration as suggested in its reference manual\footnote{https://www.tensorflow.org/install/source} on the release 2.4.0. After building the library, we simply run all Python files inside \textit{tensorflow/tensorflow/python} directory ending with \textit{*.py}. Similarly, we build PyTorch from the source using the default configuration on 1.8.0. Since building PyTorch is computationally expensive, we perform compilation in parallel with four threads. Then we simply run test cases written in Python under this directory\footnote{https://github.com/pytorch/pytorch/tree/master/test}.

\noindent \textbf{RQ1.2: Comparison against the traditional DL fuzzers}
In our comparison with the baseline fuzzers, we provide details on the number of vulnerabilities that were both detected and fixed across different releases of TensorFlow and PyTorch. To conduct this evaluation, we executed the tools on a range of releases, specifically versions 2.3.0 and 2.4.0 for TensorFlow and versions 1.7.0 and 1.8.0 for PyTorch. 
In this paper, we have considered releases 2.13.0 and 1.13.1 as the most recent versions for TensorFlow and PyTorch, respectively. It is important to note that when it comes to developer APIs, we have exclusively displayed the results for FreeFuzz since the four other fuzzers are not compatible with developer APIs. When it comes to the number of APIs tested, we maintain consistency across FreeFuzz, DeepRel, and Orion, both for TensorFlow and PyTorch, to ensure a fair comparison. This uniformity is achieved by employing the same number of APIs for all three fuzzers. This approach is adopted because all of these fuzzers rely on the MongoDB database to store their test inputs. As for DocTer, we utilize their proprietary seed database that is included in their replicated package. 

\noindent \textbf{RQ1.3: Comparison against the LLM-based DL fuzzers}
In order to compare {\tool} with LLM-based DL fuzzers, we use the same releases and reported vulnerabilities mentioned in their replication package. The rationale is the compatibility issue of LLM-based fuzzers with earlier releases of TensorFlow and PyTorch (please see subsection \ref{sec:4.3}). Hence, we use TensorFlow 2.10.0 and PyTorch 1.12.0 for the comparison.  
\begin{table}[t!]
\caption{The number of covered APIs. The number of developer APIs we collected for {\tool} is in parenthesis.} 
 \setlength{\tabcolsep}{4pt}
 \resizebox{1\columnwidth}{!}{
\begin{tabular}{lccccccc}
\toprule
& FreeFuzz & DocTer & DeepRel & TitanFuzz & AtlasFuzz & Orion & \% Improvement \\
\midrule
PyTorch    & 470  & 498 & 1071  & 1,329  & 1,377 & 1,751 & \textbf{27.1\%} \\
TensorFlow & 688  &  911 &  1,902 & 2,215  & 2,309 & 1,213 (2,824) & \textbf{74.8\%} \\
\midrule
Total     &   1,115  & 1,409 & 2,973 &  3,544 & 3,606 & 5,788  & \textbf{60.5\%} \\
\bottomrule
\end{tabular}
}
\label{tab:allapi}
\end{table}

\subsubsection{Results}

\noindent \textbf{API Coverage Results}. 
Table~\ref{tab:allapi} shows the number of APIs covered by the five fuzzers. 
Please note that the analysis is based on {\tool} and AtlasFuzz which is the state-of-the-art DL fuzzer. In the case of PyTorch, TitanFuzz covers 1,329 APIs while {\tool} covers 1,751 representing a significant 31\% improvement over AtlasFuzz. Similarly, for TensorFlow, AtlasFuzz covers 2,215 APIs while {\tool} covers 4,037 APIs which is an impressive 74.8\% improvement over AtlasFuzz. This is because {\tool} covers both end-user and developer APIs. In total, {\tool} covers 60.5\% more API compared to AtlasFuzz. 

\noindent \textbf{{\tool} vs traditional DL fuzzers}. 
Table \ref{tbl:baselineperfdeveloper} shows the detection effectiveness of {\tool} compared to FreeFuzz on developer APIs. As shown, {\tool} outperforms FreeFuzz on both TensorFlow and PyTorch releases significantly. More specifically, {\tool} detects 79 and 60 more vulnerabilities versus FreeFuzz. Table~\ref{tbl:baselineperf} shows the detection effectiveness of {\tool} compared to the baseline fuzzers. It is observable that {\tool} excels in terms of vulnerability detection for both end-user and developer APIs on TensorFlow 2.3.0, {\tool} stands out as the most effective. {\tool} takes the lead, detecting 137 vulnerabilities for end-user APIs, showcasing its robustness in vulnerability detection on TensorFlow 2.4.0. In terms of PyTorch 1.7.0, {\tool} outperforms all three tools with 5 more vulnerabilities compared to DocTer which is the second-best fuzzer on this release. This suggests that both {\tool} and DocTer are highly effective fuzzers for this release. Lastly, regarding PyTorch 1.8.0, {\tool} surpasses FreeFuzz and DocTer in vulnerability detection but is surpassed by DeepRel.

\begin{figure*}[t]
\begin{subfigure}{.3\textwidth}
  \centering
  \includegraphics[width=.8\linewidth]{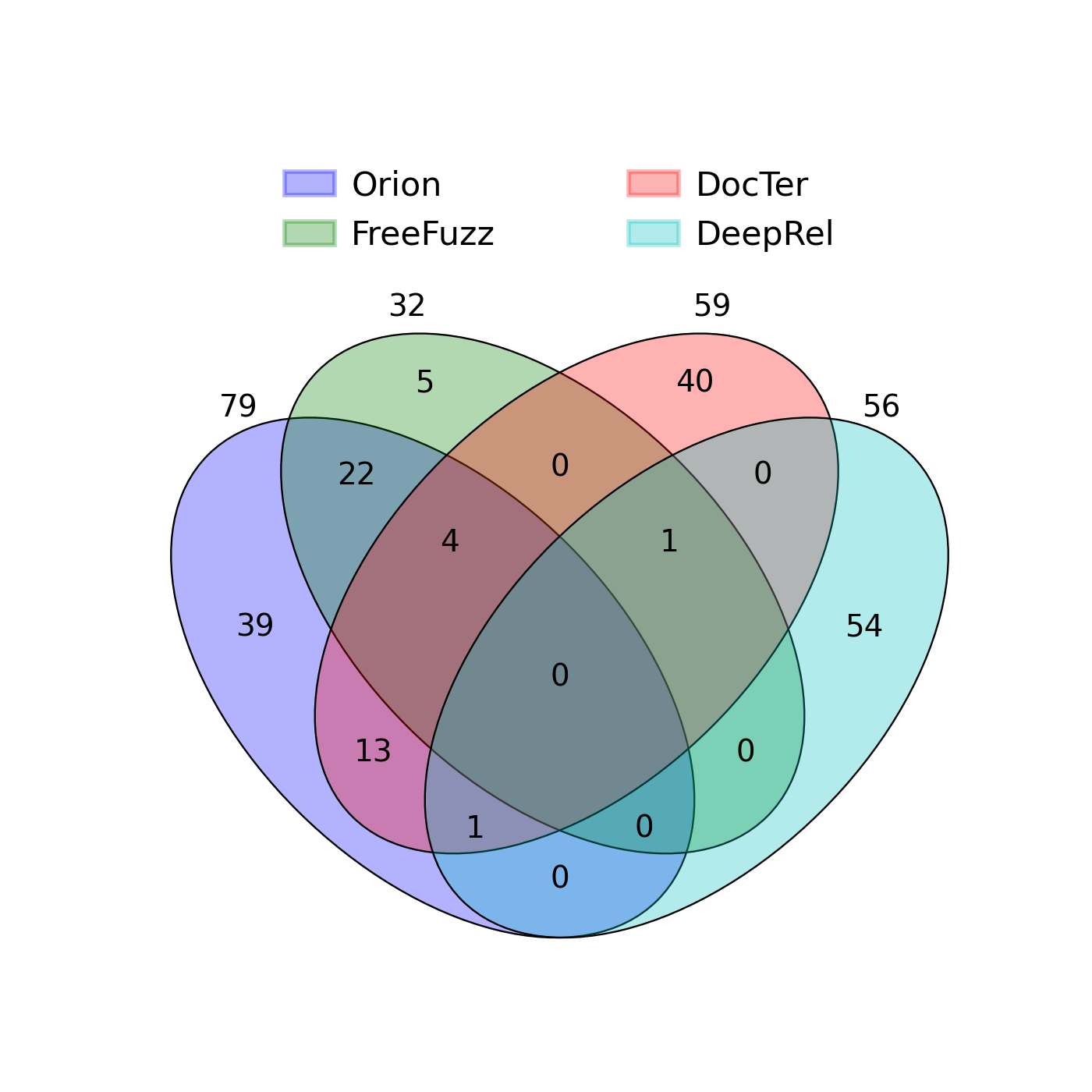}
  \caption{TensorFlow-2.3.0}
  \label{fig:overlap230}
\end{subfigure}%
\begin{subfigure}{.3\textwidth}
  \centering
  \includegraphics[width=.8\linewidth]{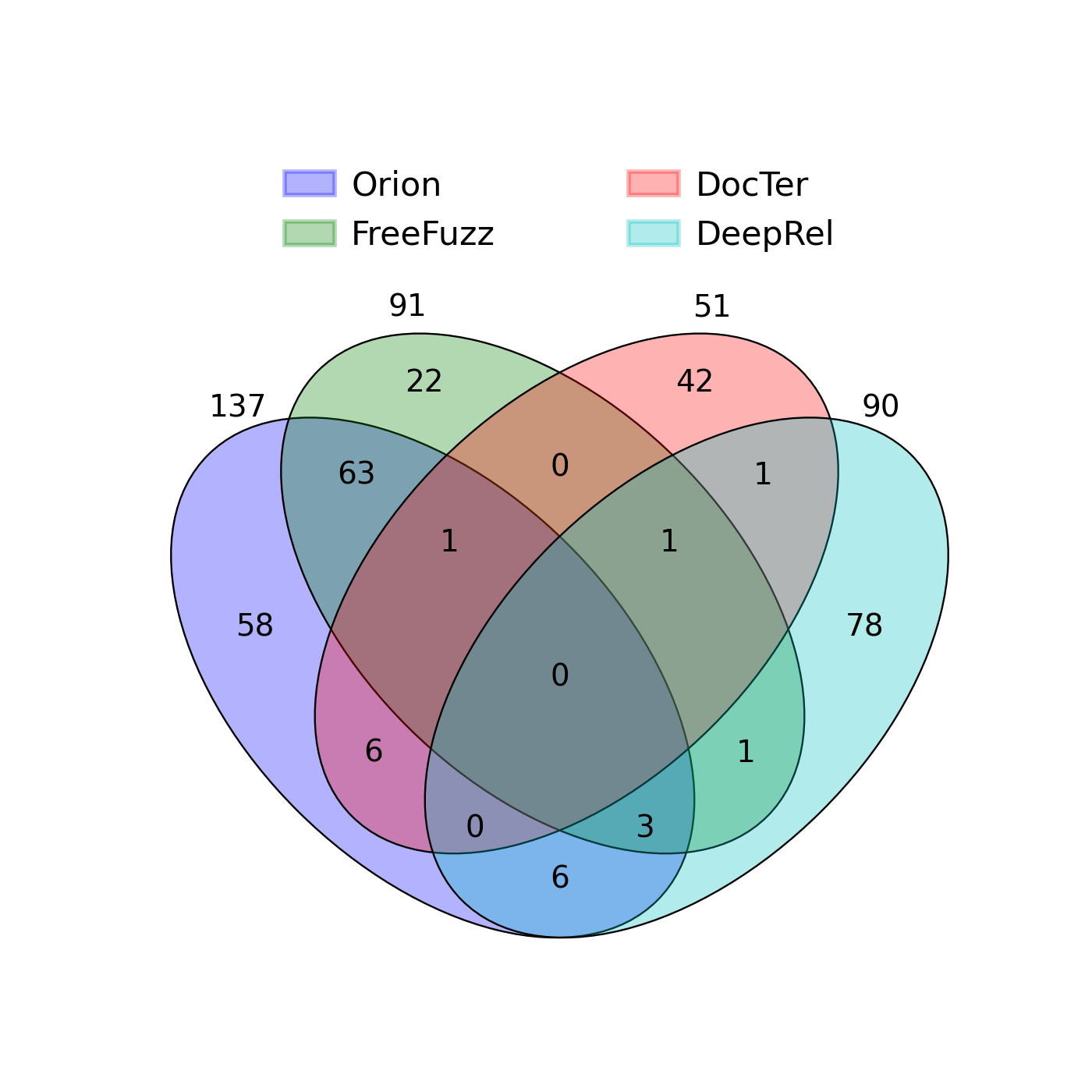}
  \caption{TensorFlow-2.4.0}
  \label{fig:overlap240}
\end{subfigure}
\begin{subfigure}{.3\textwidth}
  \centering
  \includegraphics[width=.8\linewidth]{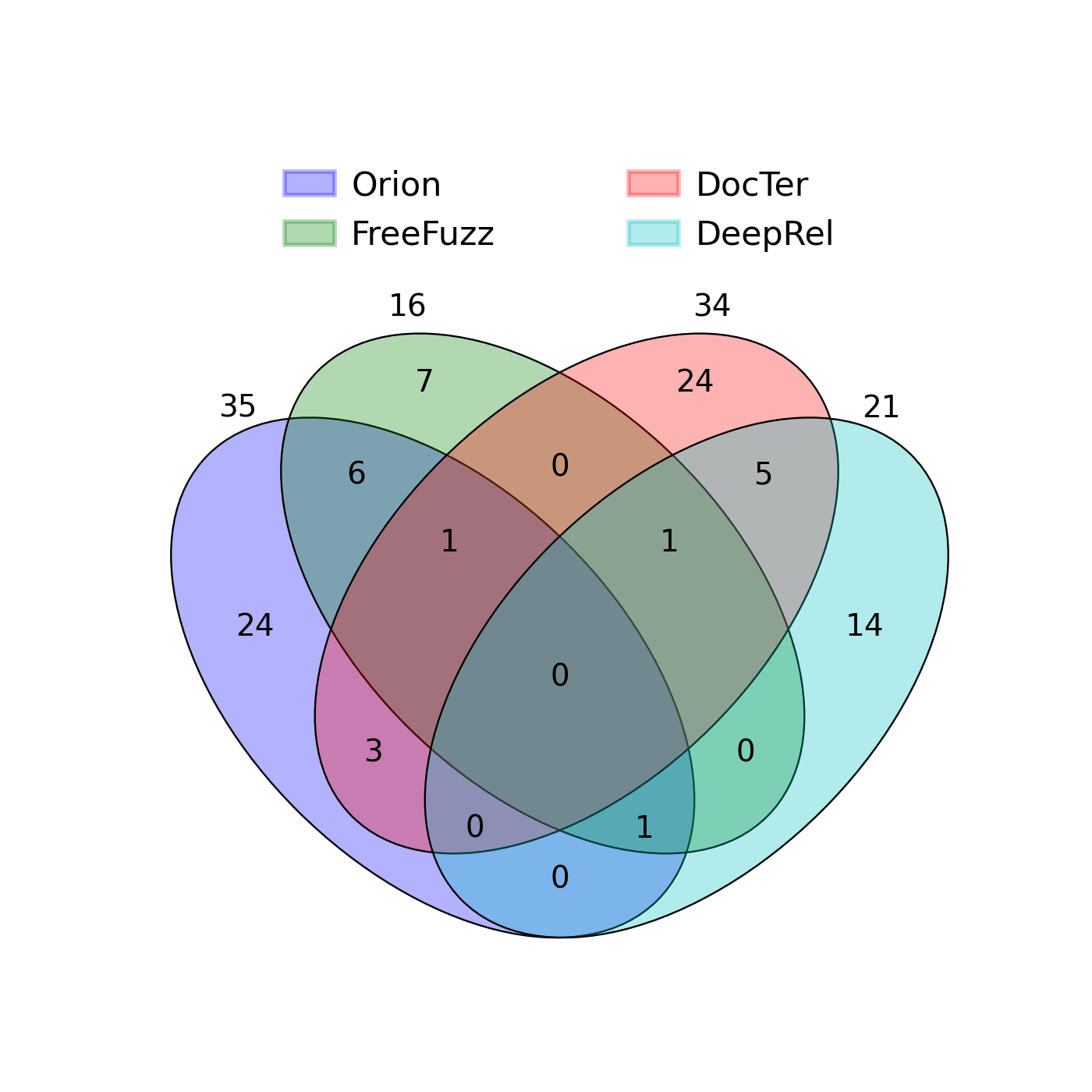}
  \caption{PyTorch-1.7.0}
  \label{fig:overlap170}
\end{subfigure}

\begin{subfigure}{.3\textwidth}
  \centering
  \includegraphics[width=.8\linewidth]{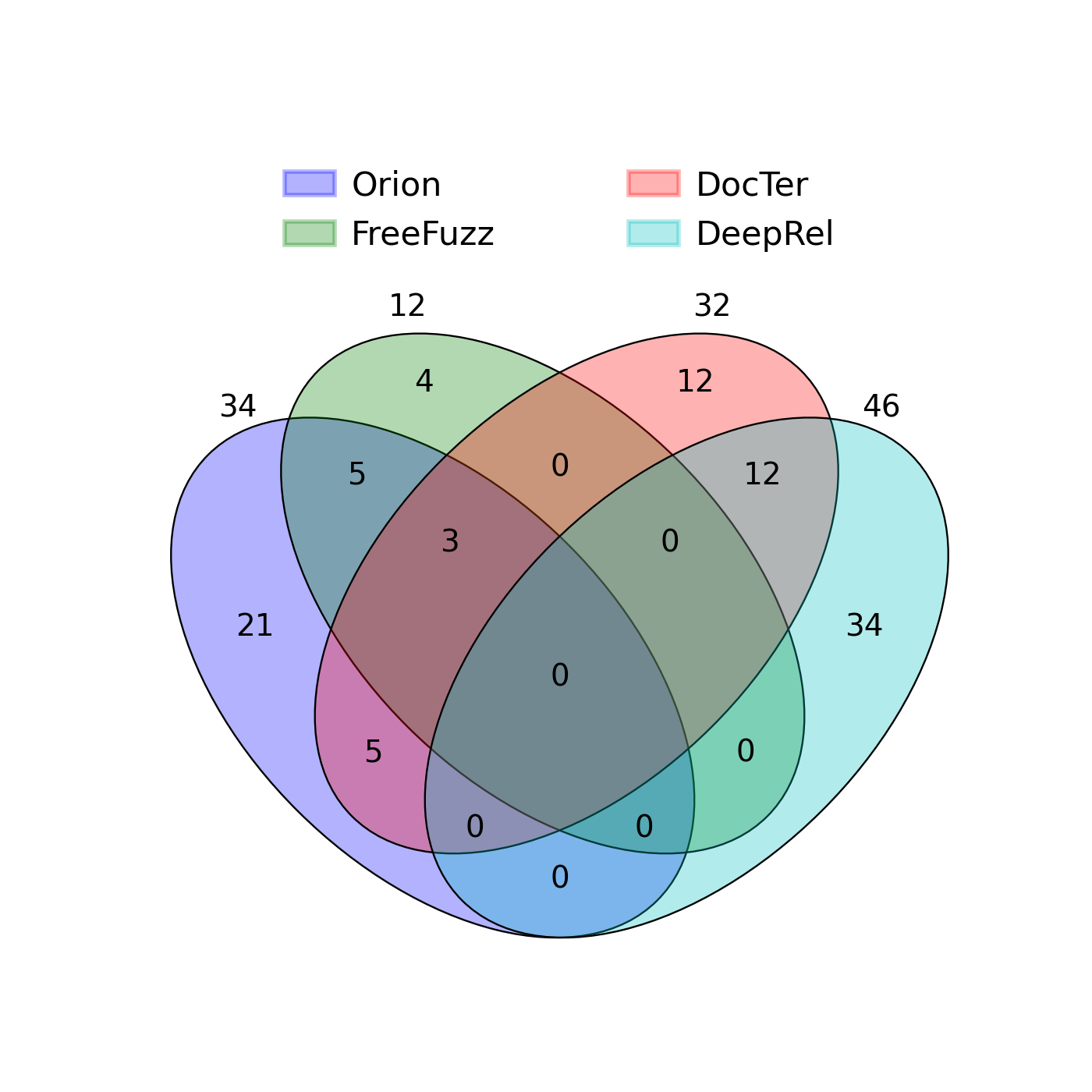}
  \caption{PyTorch-1.8.0}
  \label{fig:overlap180}
\end{subfigure}
\begin{subfigure}{.3\textwidth}
  \centering
  \includegraphics[width=.8\linewidth]{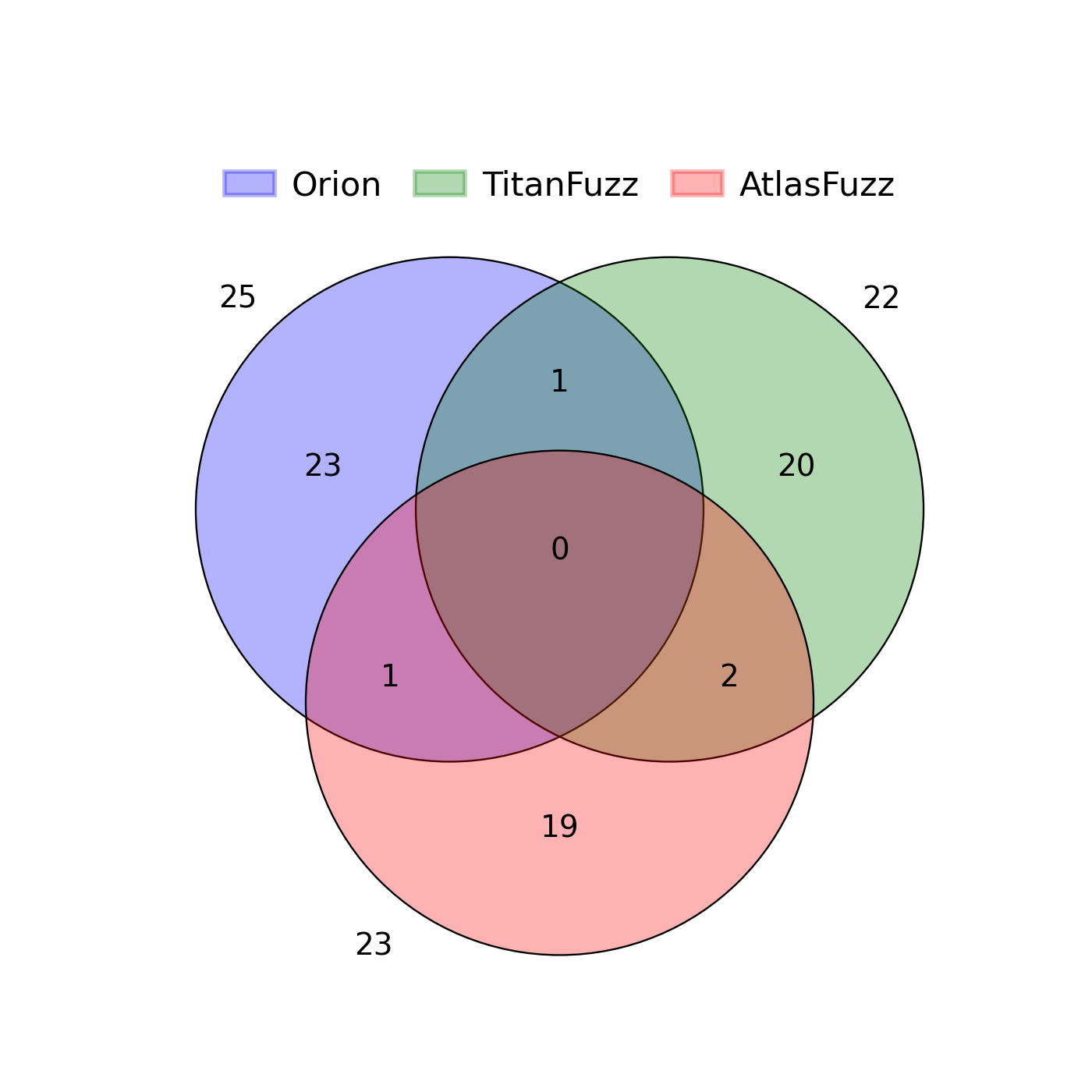}
  \caption{TensorFlow-2.10.0}
  \label{fig:tfllmoverlap}
\end{subfigure}
\begin{subfigure}{.3\textwidth}
  \centering
  \includegraphics[width=.8\linewidth]{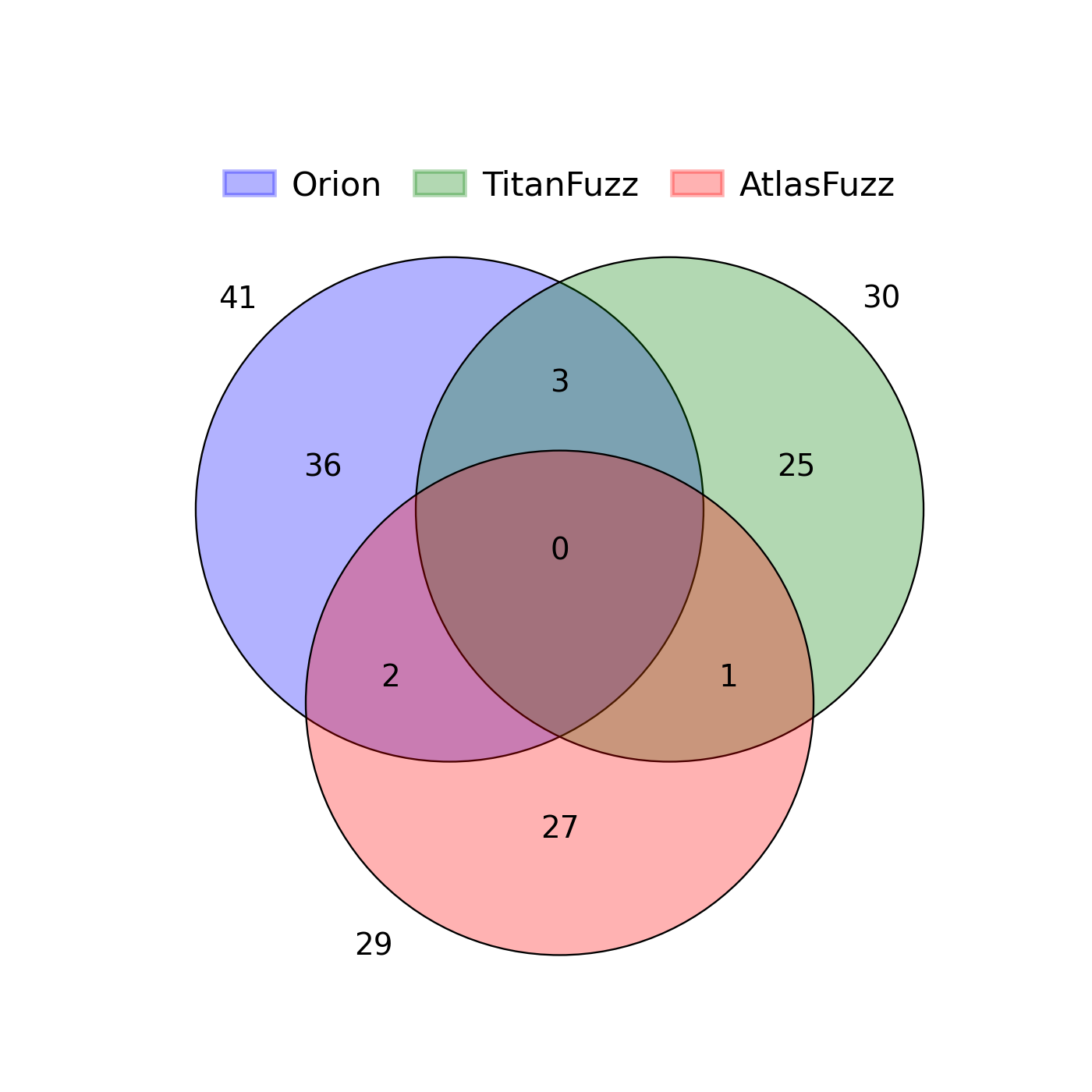}
  \caption{PyTorch-1.12.0}
  \label{fig:torchllmoverlap}
\end{subfigure}
\caption{Overlap of the number of detected bugs on TensorFlow and PyTorch releases. }
\label{fig:tfoverlaps}
\end{figure*}

Figures \ref{fig:tfoverlaps} present Venn diagrams illustrating the number of vulnerabilities detected by both {\tool} and traditional baseline methods. A key observation is that {\tool} outperforms in discovering a greater total number of vulnerabilities across both TensorFlow and PyTorch releases. Regarding traditional DL fuzzers, {\tool} consistently detects the highest number of overlapping vulnerabilities when compared to FreeFuzz across all releases of TensorFlow and PyTorch. In contrast, when compared to LLM-based fuzzers, {\tool} detects fewer overlapping vulnerabilities. This difference can be attributed to the fact that {\tool} focuses on security vulnerabilities, whereas LLM-based fuzzers primarily identify general DL bugs.

\noindent \textbf{{\tool} vs LLM-based DL fuzzers}.
Table \ref{tbl:baselinellm} offers a comprehensive comparison between {\tool} and LLM-based DL fuzzers, namely TitanFuzz and AtlasFuzz. The results demonstrate {\tool}'s superior performance, particularly evident in TensorFlow 2.10.0 and PyTorch 1.12.0, where it exhibits a substantial improvement of 13.63\% and 18.42\%, respectively. Furthermore, Figure \ref{fig:tfoverlaps} provides insights into the overlap in the number of detected vulnerabilities among {\tool}, TitanFuzz, and AtlasFuzz. It's worth noting that the total number of overlaps among the three tools is quite limited. This observation underscores the fact that each tool excels in identifying different categories of vulnerabilities. Specifically, {\tool} primarily focuses on the detection of crash vulnerabilities, while TitanFuzz and AtlasFuzz concentrate on logical vulnerabilities. This divergence in focus results in a higher intersection of detected vulnerabilities in the case of PyTorch, as illustrated in Figure~\ref{fig:torchllmoverlap}, compared to TensorFlow. In summary, these findings highlight {\tool}'s superiority over LLM-based DL fuzzers in specific versions of TensorFlow and PyTorch, emphasizing its ability to excel in the detection of crash vulnerabilities and the distinct nature of vulnerabilities detected by each tool.

 \begin{table}[t]
\caption{Comparison of Orion with FreeFuzz on developer APIs. ``Det'' denotes the number of detected bugs and ``Fixed'' is the number of fixed bugs.} 
\vspace{-0.1in}
\begin{tabular}{lcccc}
\toprule
                      & \multicolumn{4}{c}{TensorFlow}             \\
\midrule
\multirow{2}{*}{Tool} & \multicolumn{2}{c}{2.3.0} & \multicolumn{2}{c}{2.4.0}  \\
                      & Det          & Fixed       & Det          & Fixed              \\
\midrule
FreeFuzz              & 67       & 66      & 60       & 60            \\
Orion         & \textbf{146}     & \textbf{142}    & \textbf{120}     & \textbf{113}           \\
\bottomrule
\end{tabular}
\label{tbl:baselineperfdeveloper}
\end{table}

\begin{table}[t]
\caption{Comparison of Orion with the traditional fuzzers on end-user APIs.} 
\begin{tabular}{lcccccccc}
\toprule
                      & \multicolumn{4}{c}{TensorFlow}                          & \multicolumn{4}{c}{PyTorch}                             \\
\midrule
\multirow{2}{*}{Tool} & \multicolumn{2}{c}{2.3.0} & \multicolumn{2}{c}{2.4.0} & \multicolumn{2}{c}{1.7.0} & \multicolumn{2}{c}{1.8.0} \\
                      & Det          & Fixed       & Det          & Fixed       & Det         & Fixed        & Det         & Fixed        \\
\midrule
FreeFuzz              & 31       & 29      & 90      & 89    & 15          & 10           & 11          & 8            \\
DeepRel               & 68           & 8          & 111         & 23       & 20           & 17               & \textbf{75}          & \textbf{74}          \\
DocTer                & 68           & 56          & 53           & 42          & 33          & 32          & 31          & 30           \\
Orion         & \textbf{99}     & \textbf{91}    & \textbf{137}     & \textbf{124}    & \textbf{38}          & 15           & 34          & 12           \\
\bottomrule
\end{tabular}
\label{tbl:baselineperf}
\end{table}

\begin{table}[t]
\caption{Comparison of {\tool} with the  LLM-based DL fuzzers.}
\vspace{-0.1in}
\begin{tabular}{lcc}
\toprule
\multirow{1}{*}{Tool} & TensorFlow & PyTorch \\
\midrule
TitanFuzz             & 22         & 30       \\
AtlasFuzz             & 22         & 38       \\
Orion                 & 25         & 45      \\
Improvement        & \textbf{13.63\%}           & \textbf{18.42\%} \\
\bottomrule
\end{tabular}
\label{tbl:baselinellm}
\end{table}

\mybox{\textbf{Answer to RQ1:} {\tool} covers 
more APIs than the selected baselines as {\tool} considers both end-user and developer APIs. Regarding the detection performance, in general, {\tool} detects 69.3\% and 35.5\% more vulnerabilities than DeepRel on TensorFlow releases. In terms of developer APIs, {\tool} can detect 117\% and 100 more vulnerabilities compared to its extension FreeFuzz. In terms of end-user APIs, {\tool} can detect 45.58\%, 23.4\%, and 90\% on TensorFlow 2.3.0, 2.4.0, and PyTorch 1.7.0, respectively. Regarding LLM-based DL fuzzers, {\tool} can detect 13.63\% and 18.42\% more vulnerabilities on TensorFlow and PyTorch releases.}

\subsection{RQ2: Performance on Detecting New Vulnerabilities.} 
Table \ref{tbl:newbugs} presents statistics on newly detected vulnerabilities by {\tool} in the latest releases of TensorFlow and PyTorch. In total, {\tool} reports 135 vulnerabilities, 76 of which have already been confirmed by library developers, and 69 of them are new vulnerabilities. Since we report the vulnerabilities on the most recent releases for each library, at the time of writing this paper, merely 7 of them have been fixed. The number of reported vulnerabilities in TensorFlow is higher than in PyTorch since the number of tested APIs in TensorFlow is higher than in PyTorch. Also, the number of confirmed vulnerabilities within the TensorFlow library is higher since the developers of TensorFlow are more active in triaging the reported vulnerabilities. 

\begin{figure}[t!]
\begin{subfigure}{.5\textwidth}
  \centering
  \includegraphics[width=.8\linewidth]{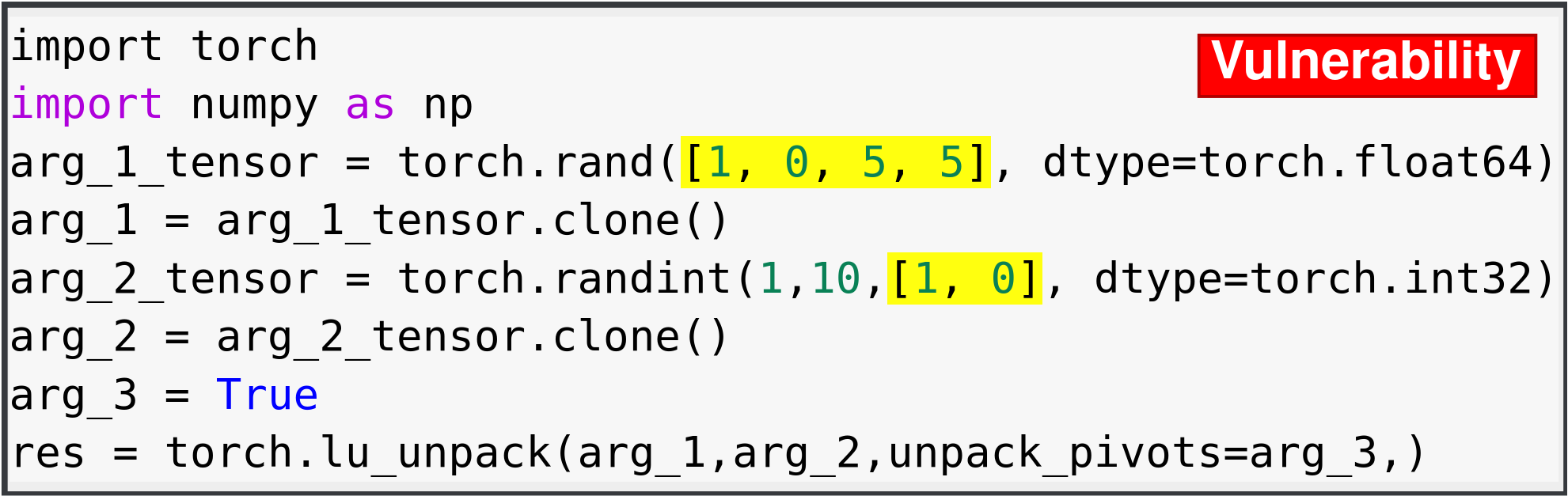}
  \caption{Segmentation fault exposed by mismatch between input tensors.}
  \label{fig:exampleBug1}
\end{subfigure}%

\begin{subfigure}{.5\textwidth}
  \centering
  \includegraphics[width=.8\linewidth]{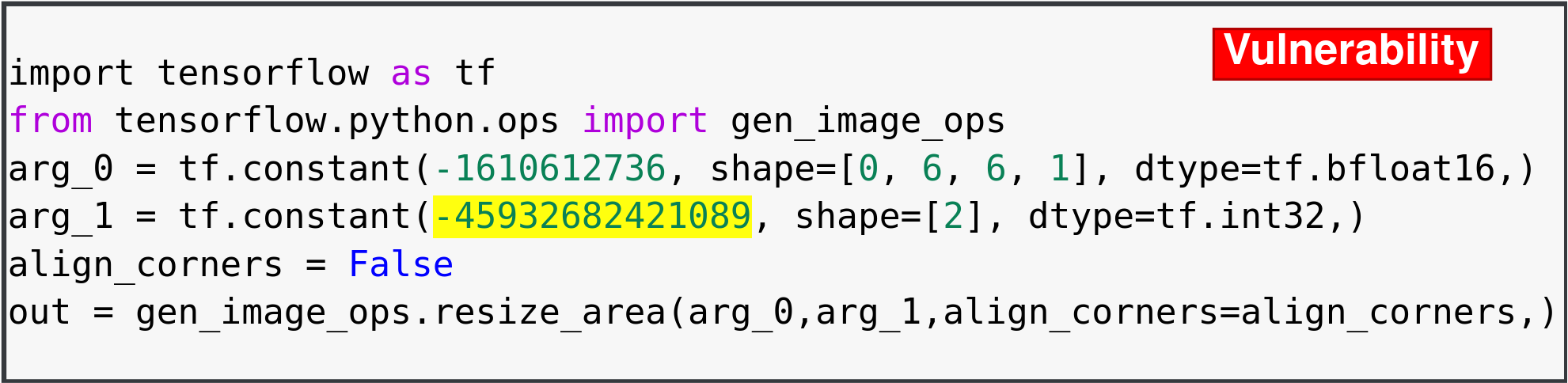}
  \caption{Segmentation fault when feeding extreme corner case tensor value.}
  \label{fig:exampleBug2}
\end{subfigure}
\caption{Example vulnerabilities detected by {\tool}.}
\label{fig:rq3bugs}
\end{figure}

Figure \ref{fig:exampleBug1} illustrates a segmentation fault example, which is one of the vulnerabilities detected by {\tool} in PyTorch. This vulnerability arises due to a misalignment in the input tensors for \textit{torch.lu\_unpack}. The highlighted syntax in the figure highlights this misalignment, as revealed by the rules employed in the \textit{Guided Input Generation} process. To put it simply, the rank of the first tensor and the second tensor should be identical. {\tool} derives its fuzzer generator from these mismatch rules, which are constructed based on historical vulnerabilities that share the same root cause.

Figure \ref{fig:exampleBug2} showcases a vulnerability that {\tool} detected in TensorFlow. This vulnerability stems from an extremely negative value within the parameter \textit{arg\_1}, leading to a segmentation fault and subsequently crashing the Python interpreter. According to the documentation, the second parameter is expected to be a \textit{1D} tensor with a precision of \textit{32 bits} for integer values. However, the documentation lacks precision in specifying that large negative integer values are not permitted. {\tool} utilizes its extreme corner case generator to detect and expose this particular vulnerability.


\begin{table}[]
\caption{The number of detected unknown, confirmed, and fixed bugs by Orion on TensorFlow 2.13.0 and PyTorch 1.13.1 as the latest releases.}
\begin{tabular}{lllll}
\toprule
Library                     & Scope                            & Total & Confirmed & New(fixed) \\
\midrule
\multirow{1}{*}{TensorFlow} & \multirow{1}{*}{Developer APIs}  & 47    & 28        & 28 (0)          \\           
                            & End-user APIs                    & 50    & 21         & 19 (0)           \\                    
                            \midrule
\multirow{1}{*}{PyTorch}    & \multirow{2}{*}{-}               & 38    & 27        & 22 (7)         \\
Total                       & -                                & 135     & 76        & 69 (7)    \\
\bottomrule
\end{tabular}
\label{tbl:newbugs}
\end{table}

\mybox{\textbf{Answer to RQ2:} 
{\tool} can detect 135 vulnerabilities, which 76 of have been confirmed by library developers. Among the confirmed vulnerabilities, 69 are unknown in the latest versions of TensorFlow and PyTorch, and 7 of them have been already fixed by library developers. The rest are pending for further confirmation.}

\subsection{RQ3: Ablation Study of Fuzzing Heuristic Rules.}


Table~\ref{tbl:rulecontrib} illustrates the contributions of fuzzing heuristic rules in the detection of vulnerabilities in TensorFlow and PyTorch. Overall, the rules categorized under \textit{Corner Case Input Generation} demonstrate a greater impact on detecting security vulnerabilities in TensorFlow and PyTorch releases compared to those under \textit{Guided Input Generation}. This trend can be attributed to the fact that corner case rules encompass a broader range of parameter types and vulnerable components within the API input specifications of DL APIs. Within the \textit{Guided Input Generation} category, the \textit{Tensor Shape Mismatch} rule exhibits the highest contribution, surpassing \textit{Tensor List-Indices Mismatch}. On the other hand, in the \textit{Corner Case Input Generation} category, the \textit{Large Integer Argument} rule emerges as the most effective in vulnerability detection within TensorFlow and PyTorch. This effectiveness is likely a result of the inherent characteristics of DL libraries, which frequently engage in high-performance numerical computations requiring constant memory allocation and deallocation in the backend. When exceedingly large integer values are provided as inputs to DL APIs, the backend often fails to validate these inputs properly. This leads to unauthorized memory access during dynamic memory allocation, ultimately resulting in a crash.

\begin{table}[t!]
\caption{Contribution of {\tool}'s fuzzing heuristics rules on detecting vulnerabilities in TensorFlow v2.13.0 and PyTorch v1.13.1.}
\resizebox{1\columnwidth}{!}{%
\begin{tabular}{llcc}
\toprule
Rule Notation & Rule Explanation          & TensorFlow         & PyTorch    \\
\midrule
Guided Input Generation  & & & \\
\midrule
 $\Lambda(T_i, T_j) = \underset{\textcolor{red}{s_l\neq s_k}}{T_i<v, \textcolor{red}{\mathbf{s_l^{n\times m}}} ,dtype>,T_j<v,\textcolor{red}{\mathbf{s_k^{n\times m}}},dtype>}$ &            Tensor Shape Mismatch            & 1                 &   4          \\
 \midrule
 $\Lambda(T_l,L_i) = \underset{\textcolor{red}{\mathbf{L_{i,j}\neq n | L_{i,j} \neq m}}}{T_l<v,\textcolor{red}{\mathbf{s_{n\times m}}} ,dtype>}, L_i=\{k_1,k_2,...,\textcolor{red}{\mathbf{k_j}}\}$  &       Tensor List-Indices Mismatch          & 3                 &   -       \\
\midrule
Corner Case Input Generation &  & & \\
\midrule
$case_x$ & Large Integer Argument                & 16                & 9           \\
$case_n$ & Negative Integer Argument             & 2                 & 1         \\
$case_x$& Zero Integer Argument                  & 1                 &  1         \\
\midrule
$case_x$ & Large List Element                    & 10                &  5           \\
$case_{noa}$& Invalid List Element                  & 2                 &  -           \\
\midrule
$case_x$ $\land$ $case_n$ & Negative Large Input Tensor           &-                  & 4 \\
$case_x$ & Large Input Tensor                    & 4                 &  1           \\
$case_n$ & Negative Input Tensor                 & 1                 & 2         \\
\midrule
$case_{none}$ & None Input Argument                   & 5                 &   -          \\
$case_{nan}$& NaN Input Argument                    & 2                 &    -         \\
$case_{noa}$ & Invalid Input String                  & 1                 &   -          \\
$case_{nmt}$& Empty Input Argument                  & 1                 &  -         \\
\midrule
- & -                               & 49                & 27          \\
\bottomrule
\end{tabular}}
\label{tbl:rulecontrib}
\end{table}

\mybox{\textbf{Answer to RQ3:} The \textit{Corner Case Input Generation} category stands out as the most effective in terms of fuzzing heuristic rules for detecting security vulnerabilities. Among these rules, the \textit{Large Integer Argument} rule emerges as the most effective, identifying a total of 26 vulnerabilities. In comparison, the \textit{Guided Input Generation} category is the second most effective, with the \textit{Tensor Shape Mismatch} rule proving to be the most impactful, uncovering a total of 5 vulnerabilities.}

\section{Threats to validity}

\noindent \textbf{Internal validity.} The internal validity mainly concerns whether the implementation of the tool is correct or not. To reduce such a threat, we did a code review in multiple rounds to make sure the rules were working as expected. Regarding the existing fuzzing frameworks, we did not modify their implementation and we use them directly to compare them with {\tool} and we use their default settings for comparison.

\noindent \textbf{External validity.} 
For this study, the external validity is on the generalizability of {\tool} on different DL libraries. To reduce this threat, we used two popular and widely used DL libraries including TensorFlow and PyTorch. Even though both of them use tensor-level operations and create dynamic computation graphs for DL training and testing, they have different back-end implementations. This diversity increases the validity of rules equipped with {\tool}. Unlike existing works which only cover end-user public APIs used only by end-users of DL libraries, we also collected dynamic execution information of developer APIs used specifically by library developers. In the future, we will extend {\tool} to more DL libraries. We also evaluate the effectiveness of {\tool} on a wide range of TensorFlow releases including the latest and earlier releases. We also assessed the effectiveness of {\tool} both on traditional and LLM-based DL fuzzers. 

\noindent \textbf{Construct validity.} 
The main concern regarding construct validity is the evaluation criteria used for the comparison of different tools or techniques. To reduce such threats, we use four metrics for the comparison, following existing work~\cite{wei2022free, xie2022docter}. In the future, more metrics will be explored. 
\section{Related work}

\subsection{Testing DL libraries}
Deep learning libraries have been widely used to assist users in the training and prediction tasks of Deep Neural Networks (DNNs) \cite{goodfellow2016deep} such as image classification~\cite{algan2021image, yu2022coca, ccalik2018cifar, wang2022git,hu2022scaling, sariyildiz2020learning, chen2022visualgpt}, natural language processing~\cite{minaee2017automatic, oh2022entropy, gessler2022microbert, tang2022improving, liu2019roberta, liu2018darts, howard2018universal, dai2015semi}, and software engineering~\cite{foreman2019vulnerability, beltramelli2018pix2code, chen2021evaluating, yin2018tranx, yasunaga2020graph}. 

Researchers have conducted numerous research studies~\cite{lemon, pham2019cradle, pei2017deepxplore, zhou2020deepbillboard, zhang2018deeproad, graese2016assessing, agarwal2022exploring, dai2022deep, choi2022argan, zhang2019adversarial, bose2020adversarial} for testing DL libraries.  One of the first steps toward testing DL libraries is the framework proposed by \cite{pham2019cradle} which is a new approach that focuses on finding and localizing bugs in deep learning (DL) software libraries. It addresses the challenge of testing DL libraries by performing cross-implementation inconsistency checking to detect bugs and leveraging anomaly propagation tracking and analysis to localize the faulty functions that cause the bugs. LEMON \cite{lemon} extends CRADLE by proposing a mutation-based framework where It designs a series of mutation rules for DL models in order to explore different invoking sequences of library code and hard-to-trigger behaviors. Another line of research is API-level testing of DL libraries \cite{xie2022docter, wei2022free, deng2022fuzzing, kang2022skipfuzz} where each API is considered a subject for fuzzing based on a set of predefined or random mutation rules. For example, DocTer \cite{xie2022docter} designed to analyze API documentation to extract deep learning (DL)-specific input constraints for DL API functions. FreeFuzz \cite{wei2022free} instrumented DL API calls from different sources and performed instrumentation to trace dynamic execution information of DL APIs. Then these instrumentations were used for random fuzzing where type and value mutations were applied. One limitation of FreeFuzz is that it does not consider the relational property of APIs that have similar names and parameter signatures. DeepRel \cite{deng2022fuzzing} further extended FreeFuzz to infer potential API relations based on API syntax and semantic information, synthesizes test programs for invoking these relational APIs, and performs fuzzing to find inconsistencies and bugs. A recently proposed fuzzer called SkipFuzz \cite{kang2022skipfuzz} uses active learning to infer the input constraints of each API function and generate valid inputs. The active learner queries a test executor for feedback, which is used to refine hypotheses about the input constraints. 

Our work differs from existing DL fuzzers in several key aspects. FreeFuzz~\cite{wei2022free} and DeepRel~\cite{deng2022fuzzing} adopt approaches that involve generating random test inputs for API-level testing of TensorFlow and PyTorch. DocTer~\cite{xie2022docter}, on the other hand, focuses on extracting a set of input generation constraints from API reference documentation. In the context of LLM-based DL fuzzers, such as TitanFuzz~\cite{deng2023large1} and AtlasFuzz~\cite{deng2023large2}, the emphasis is on leveraging LLMs to model API usage sequences and context information for API-level fuzzing which mostly expose general DL bugs, not security vulnerabilities. In contrast, our approach is centered around the construction of fuzzing heuristic rules based on historical security vulnerabilities. These rules are designed to mimic real-world corner cases test inputs when fuzzing DL APIs, thereby discovering critical vulnerabilities within the backend implementation of TensorFlow and PyTorch. Additionally, {\tool} employs guided input generation rules to model correlations among input arguments of DL APIs, addressing one of the main root causes of security vulnerabilities in DL libraries.



\section{Conclusion}
In this paper, we proposed {\tool}, the first step toward building a semi-automated fuzzing framework based on a set of fuzzing heuristics rules built on top of the history security vulnerabilities in TensorFlow and PyTorch. More specifically, {\tool} performs test input generation for fuzzing via mining historical data from open source including API documentation, public repositories in GitHub, and library tests. We built the fuzzing heuristics rules based on 33 unique root causes of security vulnerabilities. We extensively evaluated {\tool} versus three traditional DL fuzzers, including FreeFuzz, DeepRel, and DocTer as well as two state-of-the-art LLM-based fuzzers on more than 5k end-user and developer DL APIs. {\tool} reports 135 vulnerabilities 76 of which are confirmed by the community of DL library developers. Among the 76 confirmed vulnerabilities, 69 are new vulnerabilities, 7 of them have been fixed after we reported them and the left are awaiting confirmation. In terms of end-user APIs and compared to most recent traditional DL fuzzers, i.e., DeepRel, {\tool} detects 31.8\% and 90\% more vulnerabilities respectively. In comparison with the state-of-the-art LLM-based fuzzer, e.g., AtlasFuzz, {\tool} detects 13.63\% and 18.42\% more vulnerabilities on TensorFlow and PyTorch.
\section{Data availability}

We share the source code of {\tool}, bug list, and data at~\cite{ourdata}.

\bibliographystyle{ACM-Reference-Format}
\bibliography{paper}


\end{document}